\documentstyle[11pt]{article}
\input{epsf}
\begin{document}
\title{\textsl{Prefect Transfer of Quantum States on Spin
Chain with Dzyaloshinskii- Moriya interaction in inhomogeneous
Magnetic field}}
\author{S. Salimi
  \thanks{Corresponding author: E-mail: shsalimi@uok.ac.ir} , \
B. Ghavami
  \thanks{E-mail: badie.ghavami@gmail.com} \ and \
  A. Sorouri
  \thanks{E-mail:a.sorouri@uok.ac.ir}
 \\ {\small Department of Physics,
University of Kurdistan, Sanandaj 51664, Iran.} }  \pagebreak
\date{\textit{\today}}
%%%%%%%%%%%%%%%%%%%%%%%%%%%%%%%%%%%
\maketitle \abstract{In
 this work, we use the Hamiltonian of a modified
Dzyaloshinskii-Moriya model and investigate the perfect transfer of
 the quantum state on the spin networks. In this paper, we calculate fidelity in which fidelity  depends on magnetic
field and another parameters. Then, by using the numerical analysis we show that the fidelity of the transferred state is determined by magnetic field $B$,
exchange coupling $J$ and the Dzyaloshinskii- Moriya
interaction  $D$. We also found that the perfect transfer
 of the quantum state is possible with condition  $B \gg \Gamma^2\omega^{N/2}$ where $\Gamma =((J+iD)/2)$ and $\omega=\Gamma^*/ \Gamma$.}\\
\noindent {\bf Keywords}: $XY$ Spin-Chain, Teleportation,
Dzyaloshinskii-Moriya interaction.
 \\
%\twocolumn
%\newpage
%%%%%%%%%%%%%%%%%%%%%%%%%%%%%%%%%%%%%%%%%%%introduction$$$$$$$$$$$$$$$$$$$$$$
%\noindent {\bf\textit{ Introduction}}\\
\newpage
\section{Introduction}

Quantum communication is the act of transferring a quantum state
from one place to another. By far, its most well known application
is quantum key distribution through which a secret random key can
be established between distant parties with its security
guaranteed by quantum mechanics [1, 2]. Jiankui He, Qing Chen,
Lei Ding, Shao-Long Wan have been investagated  the quantum
state transfer in randomly coupled spin chains [3]. By using local
memories storing the information and dividing the process into
transfer and decoding, conclusive transfer is spontaneously
achieved with just one single spin chain. X.Q. Xi, J.B.Gong, T.
Zhang, R.H.Yue, and W.M. Liu [4] have investigated  simple
and explicit designs of short isotropic "XY" spin chains for
perfect quantum
state transfer are obtained analytically.

Jing-Ling Chen and Qing-Liang Wang \cite{reff} have shown that
perfect state transfer over arbitrary distances is possible for
a simple unmodulated spin chain by some schemes. The transfer of
a single qubit state has been investigated in detail by
Christandl et al. [Phys. Rev. Lett. 92, 187902(2004)] through a
modified Heisenberg XX model Hamiltonian$ H_G$. The previous study of
Christandl was restricted to the excitation states of $H_G$
(i.e., which correspond to the second subspace of the Hilbert
space of $H_G$). They extended their study to the case of the
high-excitation states, and  the entangled states in the
 form, $|\psi\rangle=\alpha|00...0\rangle+\beta|11...1\rangle$
can be perfectly transferred on the spin chain.

In quantum information processing, transferring a quantum state
from one qubit to another is necessary in many cases, e.g.,
quantum key distribution \cite{5}  quantum teleportation \cite{6}, and
quantum computation \cite{7}. For long-distance quantum communication,
there seems no doubt that photons should be the information
carriers. For short-distance quantum communication and especially
in a solid-state environment, much of the ongoing research
efforts are being devoted to spin chains as promising "quantum
wires". Experimental studies of short spin chains using nuclear
magnetic resonances have also emerged \cite{8,9} Bose first proposed
to use a spin chain as a quantum channel for quantum state
transfer between different qubits located on two ends of a spin
chain \cite{10}. Bose's scheme has motivated many studies focusing on
the possibility of perfect quantum state transfer in spin systems
[12- 26], i.e., transferring a quantum state with a
fidelity one.

In this work, we use the Hamiltonian of a modified
Dzyaloshinskii-Moriya model and investigate the perfect transfer of
 the quantum state on the spin networks. In this paper, we calculate fidelity in which  depends on magnetic
field, Dzyaloshinskii-Moriya interaction and another parameters. Then, by using the numerical analysis we show that the perfect transfer
 of the quantum state is possible with condition  $B \gg \Gamma^2\omega^{N/2}$.

\section{Perfect quantum state transfer}
Let us consider the Hamiltonian of a modified
Dzyaloshinskii-Moriya model
\begin{equation}\label{Eq:1}
H=-\sum_{j=1}^{N}(J(\sigma_{j}^{x}\sigma_{j+1}^{x}+
\sigma_{j}^{y}\sigma_{j+1}^{y})
+D(\sigma_{j}^{x}\sigma_{j+1}^{y}-\sigma_{j}^{y}
\sigma_{j+1}^{x})+\frac{B}{2}(\sigma_{j}^{z}+1)),
\end{equation}
where $J$ is coupling strength between lattices $j$ and $j+1$,
$D$ is the $z$ component of the Dzyaloshinskii- Moriya
interaction, $\sigma^{x,y,z}$ are the Pauli matrices, $N$ is the
number of sites and $B$ is the strength of the external
magnetic field on every site.

Now we define the raising and lowering operators as
$\sigma^{+}=\sigma^x+i\sigma^y$ and
$\sigma^-=\sigma^x-i\sigma^y$. Therefore, the Hamiltonian(1{\label{Eq:1}}) can be transformed into
\begin{equation}\label{Eq:2}
H=-\sum_{j=1}^{N}(\Gamma_{j,j+1}\sigma_j^{+}\sigma_{j+1}^-
+\Gamma_{j,j+1}^{*}\sigma_j^-\sigma_{j+1}^{+}
+\frac{B}{2}(\sigma_j^z+1))
\end{equation}
where $\Gamma_{j,j+1}=((J+iD)/2)$. The Hamiltonian
$H$ obviously describes a nearest-neighbor interaction spin chain.
Hamiltonian has $2^N$ complete and orthogonal eigenvectors, which produce
the Hilbert space of spin chain $\mathcal{H}$. The Hilbert space of $\mathcal{H}$ can be
divided into $N+1$ subspaces based on the population of reversed
spin [1]. The first subspace has only one eigenvector with
zero-value eigenvalue, i.e.
\begin{equation}
|\psi_0\rangle=|00...0\rangle,\hspace{5mm}
H|\psi_0\rangle=E_0|\psi_0\rangle,\hspace{5mm} E_0=0,
\end{equation}
where we have denoted $|0\rangle$ as the state of spin-down
$|-\rangle$, and $|1\rangle$ as the state of spin-up $|+\rangle$.
The ground state $|\psi_0\rangle$ is a state with all spin down.
The first- excitation states, contains $N$ states, which have the
following forms
\begin{equation}\label{4}
|\psi_1\rangle^{(k)}=\sum_{m=1}^Na_k(m)\phi(m),\hspace{5mm}H|\psi_1\rangle^{(k)}=
E^{(k)}_1|\psi_1\rangle^{(k)},\hspace{5mm}k=1,2,...,N,
\end{equation}
where
\begin{equation}
\phi(m)=|00...1_m...0\rangle.
\end{equation}
Representing Hamiltonian in $\phi(m)$ basis, we have
\begin{equation}
H=-\pmatrix{B&\Gamma&0&0&...&0&0\cr\Gamma^{*}&
B&\Gamma&0&...&0&0\cr 0&\Gamma^{*}&B&\Gamma&...&0&0
\cr\vdots&\vdots&\vdots&\vdots&\ddots\cr
0&0&0&0&\ldots&\Gamma^{*}&B}_{N\times N},
\end{equation}
where $\Gamma=\Gamma_{j,j+1}$ and  $B \gg \Gamma^2\omega^{N/2}$. The eigenvalues and eigenvectors for
the above Hamiltonian $H$ obtained as
\begin{equation}
E(\theta)=-(B+2\sqrt{\Gamma^{*} \Gamma
}\cos(\theta))\ ; \quad\quad  -\pi\leq\theta\leq\pi,
\end{equation}
 and
\begin{equation}\label{8}
|\theta\rangle=\beta\pmatrix{\sin(\theta)\cr
\omega^{1/2}\sin(2\theta)\cr \omega\sin(3\theta)\cr
\omega^{3/2}\sin(4\theta)\cr
\omega^{2}\sin(5\theta)\cr\cr\vdots\cr\cr
\omega^{(N-1)/2}\sin(N\theta)+\frac{\Gamma^{*}}{B}\omega^{(N-2)/2}\sin((N+1)\theta)
}_{N\times1},
\end{equation}
respectively, where $\omega=\Gamma^*/ \Gamma$ and $\beta$ is normalization
coefficient which is obtained from
\begin{equation}\label{Eq:9}
\beta^2=1/\{(\omega^{(N-1)}(\sin(N\theta)+\frac{\sqrt{\Gamma^*\Gamma}}
{B}\sin((N+1)\theta))^2+\sum^{N-1}_{n=1}\omega^{n-1}\sin^2(n\theta))\}.
\end{equation}
We assume that the sender, Alice, has full access to the first qubit $A$ in the state $a|0\rangle+ b|1\rangle$
and that the receiver, Bab, has full access to the last qubit $B$ of chain. Then, the state of chain is
\begin{equation}
a|00...0\rangle+b|10...0\rangle=a|\underline{0}\rangle+b|\underline{1}\rangle,
\end{equation}
where $|\underline{n}\rangle$, corresponding to  $|00...1_m...0\rangle$. The coefficient of $a$ does not change with time as $|\underline{0}\rangle$
is eigenstate of $H$ with eigenvalue zero. Therefore, the state $|\underline{1}\rangle$ will evolve into the superposition of the  states with one exactly  spin up and all other  down. The time evolution of the initial state is
\begin{equation}
a|0\rangle+ b|1\rangle \rightarrow a|0\rangle+ \sum_{n=1}^{N}b_n(t)|\underline{n}\rangle.
\end{equation}

To see how much the transferred quantum state is similar to the original quantum state, a quantity called fidelity is introduced.
According to the definition, the magnitude of the fidelity is a real number between $0$ and $1$. When it is $0$, both states are completely
different, meaning that the original information is completely destroyed during the transmission process. The value $1$ for
the magnitude of the fidelity in the process shows that the perfect transfer occurs and original information is completely transmitted.

For the above-mentioned system the fidelity is  $F(t)=\langle \underline{N}|e^{-i\alpha Ht}|\underline{1}\rangle=\langle
\underline{N}|Ue^{-i\alpha H_{dig}t}U^{\dag}|\underline{1}\rangle $, where $|\underline{1}\rangle=|1000...0\rangle$ and $|
\underline{N}\rangle=|000...01\rangle$ and matrix $U$
is a unitary matrix which makes  Hamiltonian $H$ diagonalized through a similarity transformation.
Regarding   Eqs.(\ref{4}) and (\ref{8}) one can see that
$|\psi_1^{(k)}\rangle=|\theta\rangle$, $a_{\theta}(1)=\beta
\sin(\theta)$, $a_{\theta}(2)=(\omega)^{1/2}\beta \sin(2\theta)$,
... and $a_{\theta}(N)=\beta ((\omega)^{(N-1)/2}\sin(N\theta)+
\frac{\Gamma^{*}}{B}(\omega)^{(N-2)/2}\sin((N+1)\theta))$. Therefore, the fidelity for
 the first-excitation states is
\begin{equation}\label{Eq:10}
F(t)=\frac{1}{2\pi}\int_{-\pi}^{\pi}d\theta
a_{\theta}^{*}(1)a_{\theta}(N)e^{-i\alpha t E(\theta)},
\end{equation}
where $E(\theta), a_{\theta}(1)$ and $ a_{\theta}(N)$ were defined above. By substituting, one obtain
$$
F(t)=(\omega)^{(N-1)/2}\frac{e^{i\alpha Bt}}{2\pi}\times
$$
\begin{equation}\label{Eq:11}
\int_{-\pi}^{\pi}\beta^2\sin(\theta)(\sin(N\theta)+
\frac{\sqrt{\Gamma^{*}\Gamma}}{B}\sin((N+1)\theta))e^{2i\alpha
t \sqrt{\Gamma^*\Gamma}cos(\theta)}d\theta.
\end{equation}
To see the quality of transmitting the information for large $N$ (long distance) in the system, we calculate the fidelity
when $N\gg 1$. The result is illustrated in Fig.1 which it show fidelity as a functions of $t$ and $D$
which it is $0\leq F\leq 1$. Also, Fig.2 show the  changes of the fidelity in the time $t$ for constant other parameters.
\\
\begin{figure}
\label{fig:AA} \includegraphics{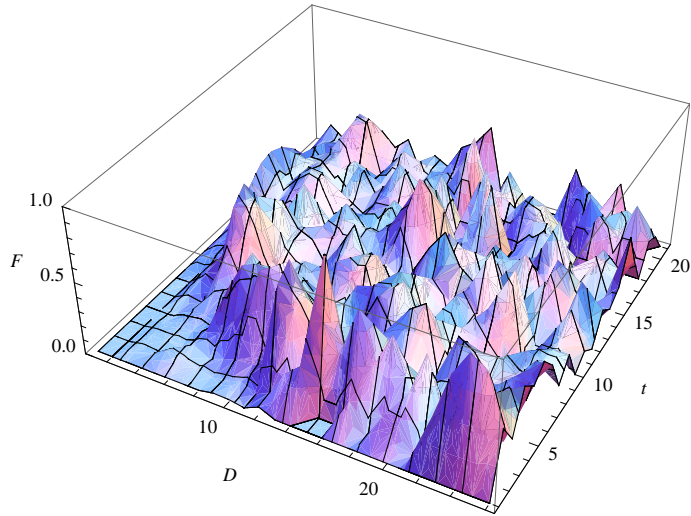}\vspace{6cm}\caption{$F$ as a function of
$t$ and $D$ where , $B=500$, $\alpha=1$, $J=1$, $N=150$.}
\end{figure}
\\

\begin{figure}
\label{fig:a1} \includegraphics{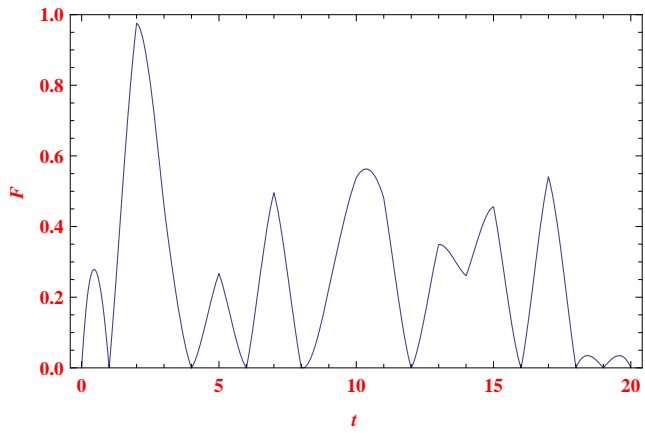}\vspace{6cm}\caption{$F$ as a function of
$t$ where , $B=500$, $\alpha=1$, $D=14.455$, $J=1$, $N=150$.}
\end{figure}
\section{Conclusion}

In this paper, we have studied the prefect transfer quantum states on spin chain. In this case, we address the problem of arranging $N$
interacting qubit, in presence of magnetic field and  Dzyaloshinskii-Moriya interaction, in a network which allows the prefect transfer of any quantum state over the longest possible distance. Then, we shown that the prefect  quantum state transfer accomplish with condition  $B \gg \Gamma^2\omega^{N/2}$.


\newpage
\begin{thebibliography}{99}
\bibitem{1}
 C.H. Bennett and G. Brassard, "Quantum Cryptography:
Public Key Distribution and Coin Tossing", Proceedings of IEEE
International Conference on Computers Systems and Signal
Processing, Bangalore India, December 1984, pp 175-179.
 \bibitem{2} A. K. Ekert, Phys. Rev. Lett. 67, 661 (1991).
\bibitem{3} Jiankui H, Qing Chen, Lei Ding, Shao-Long Wan,
Physics Letters A 372,185 - 190, (2008).
\bibitem{4} X.Q. Xi, J.B.Gong, T.
Zhang, R.H.Yue, and W.M. Liu, Eur. Phys. J. D 50, 193- 199 (2008).
\bibitem{reff} J-L. Chen and Q-l. Wang, Int. J. Ther. Phys. 46, 614 (2007).
\bibitem{5} N. Gisin et al., Rev. Mod. Phys. 74, 145 (2002).
\bibitem{6} C.H. Bennett et al., Phys. Rev. Lett. 70, 1895 (1993).
\bibitem{7} C.H. Bennett, D.P. Di Vincenzo, Nature 404, 247
(2000).
\bibitem{8} J.F. Zhang et al., Phys. Rev. A 72, 012331 (2005).
\bibitem{9} P. Cappellaro, C. Ramanathan, D.C. Cory, Phys. Rev. A
76, 032317 (2007).
\bibitem{10} S. Bose, Phys. Rev. Lett. 91, 207901 (2003).
\bibitem{11} Y. Li et al., Phys. Rev. A 71, 022301 (2005).
\bibitem{12} M. Christandl et al., Phys. Rev. Lett. 92, 187902 (2004).
\bibitem{13} F. Verstraete, M.A. Martin-Delgado, J.I. Cirac, Phys. Rev.
Lett. 92, 087201 (2004).
\bibitem{14} T.J. Osborne, N. Linden,
Phys. Rev. A 69, 052315 (2004)
\bibitem{15} D. Burgarth, S. Bose, Phys. Rev. A
71, 052315 (2005).
\bibitem{16} C. Albanese et al., Phys. Rev. Lett. 93,
230502 (2004).
\bibitem{17} T. Shi et al., Phys. Rev. A 71, 032309 (2005).
\bibitem{18} M.H. Yung, S. Bose, Phys. Rev. A 71, 032310 (2005).
\bibitem{19} P. Karbach, J. Stolze, Phys. Rev. A 72, 030301(R)
(2005).
\bibitem{20} A.
Wojcik et al., Phys. Rev. A 72, 034303 (2005).
\bibitem{21} J.B.
Gong, P. Brumer, Phys. Rev. A 75, 032331 (2007).
\bibitem{22} V. Giovannetti, D. Burgarth, Phys. Rev. Lett. 96, 030501 (2006).
\bibitem{23} C. Facer, J. Twamley, J.D. Cresser, Phys. Rev. A 77, 012334
(2008).
\bibitem{24} D.L. Feder, Phys. Rev. Lett. 97, 180502 (2006)
\bibitem{25} F.W. Strauch, C.J. Williams, Phys. Rev. B 78, 094516 (2008)
\end{thebibliography}
\end{document}